# Evaluation of the Sensitivity of RRAM Cells to Optical Fault Injection Attacks

Dmytro Petryk[1], Zoya Dyka[1], Eduardo Perez[1], Mamathamba Kalishettyhalli Mahadevaiaha[1], Ievgen Kabin[1], Christian Wenger[1,2] and Peter Langendörfer[1,2]

[1]*IHP – Leibniz-Institut für innovative Mikroelektronik,* Frankfurt (Oder), Germany
[2]*BTU Cottbus-Senftenberg,* Cottbus, Germany
petryk@ihp-microelectronics.com

*Abstract* — **Resistive Random Access Memory (RRAM) is a type of Non-Volatile Memory (NVM). In this paper we investigate the sensitivity of the TiN/Ti/Al:HfO$_2$/TiN-based 1T-1R RRAM cells implemented in a 250 nm CMOS IHP technology to the laser irradiation in detail. Experimental results show the feasibility to influence the state of the cells under laser irradiation, i.e. successful optical Fault Injection. We focus on the selection of the parameters of the laser station and their influence on the success of optical Fault Injections.**

## I. Introduction

The requirements to NVM characteristics are high density, low cost, fast write and read access using a low energy operation [1]. There is NVM based on electrically switchable resistance such as RRAM that complies with these requirements. Nowadays the investigations in the area of RRAM based devices are very active and even increasing due to the benefits that RRAM can offer. Due to its high energy efficiency RRAM is very attractive for low-power devices in the Internet of Things (IoT). The IoT consists of a lot of devices such as embedded systems, wireless sensor nodes, control systems, etc. Security is paramount for these devices but usually they do not possess sophisticated protection means. The issue is that they can be stolen to be attacked in a laboratory. One of the ways to breach the protection of these devices is to perform Fault Injection (FI) attacks.

In this paper we investigate the sensitivity of IHP 1T-1R RRAM cells based on a TiN/Ti/Al:HfO$_2$/TiN stack to optical Fault Injection attacks. The paper is structured as follows. Section II describes the phenomena of resistive switching, the structure of the attacked IHP RRAM chips and the measurement equipment to operate an IHP RRAM chip. Section III describes the equipment used to perform the optical Fault Injection attacks. Section IV presents the preliminary preparations of the attacked chips before optical Fault Injection experiments. Section V shows the results of optical Fault Injection attacks on IHP RRAM chips. Section VI concludes this work.

## II. Background Basics: RRAM

### A. Resistive switching

Resistive Random Access Memory (RRAM) is a non-volatile memory, also known as memristor that is based on the phenomena of resistive switching [2]. Resistive switching (RS) is the physical phenomenon that uses non-volatile and reversible changes of the resistance due to the application of electric stress, typically voltage or current pulsing [1], [3]. Typical RS systems are capacitor like devices, where the electrodes are a metal and the dielectric is a transition metal oxide (TMO) with insulator properties. Resistive switching in TMO with insulator property allows to use a Metal-Insulator-Metal (MIM) structure as a memory element.

### B. Physical processes at the Ti/Al:HfO$_2$ interface

There is a large diversity of physical phenomena that may lead to resistive switching. In this work we describe only one class that exploits electrochemical effects, particularly valence change mechanism (VCM). Details about other classes can be found in [1]. These electrochemical effects relate to the oxidation-reduction (redox) processes in the MIM structure, particularly in TMOs. The resistance of TMOs depends on certain oxygen stoichiometry. Redox processes in the case of VCM are provoked by an electrical voltage. This valence change is associated with a migration of oxygen vacancies in TMOs e.g. HfO$_2$. This change of the stoichiometry leads to a redox reaction that subsequently affects the conductivity of the material. Oxidation or reduction processes can be determined by the applied voltage polarity. Commonly the switching mechanism is described by the defects in a material, particularly the Frenkel defect [4]. This defect is a point defect in a crystal lattice when an atom leaves the lattice site and places itself in an interstitial position. This creates a vacancy in the lattice site. The ion that leaves the lattice site and the formed vacancy are called a Frenkel pair. In case of the HfO$_2$ layer Frenkel defects create the oxygen anions and the oxygen vacancies with charge states of 2- and 2+ respectively. Depletion or enrichment of the vacancies in a material changes the valence, i.e. affects the material conductivity. In the following we focus on describing the conductive filament (CF) forming/rupture processes that occur in the IHP MIM structure that is built on a TiN/Ti/Al:HfO$_2$/TiN stack. According to the fabrication process the growth direction of the CF is from the anode (TiN top electrode) to the cathode (TiN bottom electrode), see **Fig. 1**–*(b)*. **Fig. 1** shows the migration of Frenkel pairs in the TiN/Ti/Al:HfO$_2$/TiN stack. The details about the influence of the fabrication process on the CF growth direction can be found in [4]. The Ti layer acts as a scavenging layer for the oxygen atoms. It is used to increase the performance and reliability in RRAM cells [5].

After manufacturing of the IHP RRAM cell the Al-doped HfO$_2$ layer has a gradient of vacancies that are concentrated closely to the Ti layer at the Ti/Al:HfO$_2$ interface, i.e. Frenkel pairs are created, see **Fig. 1**–*(a)*. Under a direct electric field the oxygen anions migrate in the Ti layer and accumulate there with a higher density at the Ti/Al:HfO$_2$ interface. The migration of oxygen anions into the Ti layer can be associated with the reduction process. Reciprocally the formed oxygen vacancies inside the Al-doped HfO$_2$ layer are associated with an oxidation process. Simultaneously with the migration of oxygen anions the oxygen vacancies migrate inside the Al-doped HfO$_2$ layer to the TiN bottom electrode gathering there. As a consequence the oxygen vacancies create a CF. The CF in TMO establish the connection between anode and cathode for the current flow, see **Fig. 1**–*(b)*. It is the state with a high conductivity level, i.e. it is a Low Resistive State (LRS). When applying a reverse electric field the oxygen anions at the Ti/Al:HfO$_2$ interface may recombine with the oxygen vacancies, i.e. the reduction of the conductive Hf in the Al-doped HfO$_2$ layer starts. Simultaneously the Ti layer is oxidized by the loss of the oxygen anions. As a consequence the CF is ruptured, see **Fig. 1**–*(c)*. Rupture of the CF disconnects the anode from the cathode that significantly reduces the current through the TMO. It is the state with a low conductivity level, i.e. it is a High Resistive State (HRS). Note that a formation of CFs is not a completely predefined process due to the randomness of defects in the crystal lattice. Also the interaction of Ti and Al-doped HfO$_2$ layers at the Ti/Al:HfO$_2$ interface has a non-stoichiometric behavior. Due to these facts the thickness of CF may vary from cell to cell. It causes significant changes in the conductivity when applying equal voltages to the cells.

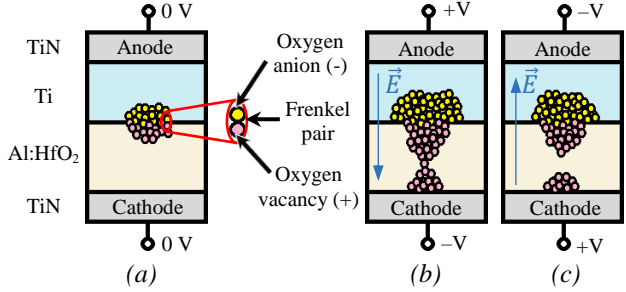
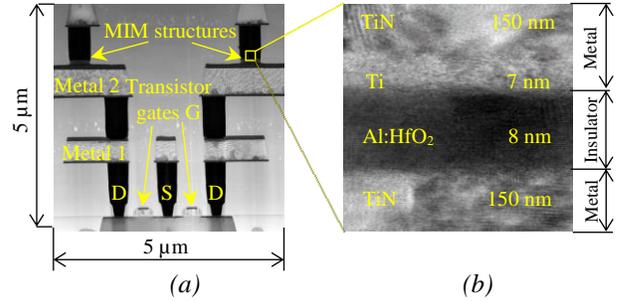

**Fig. 1.** Illustration of the MIM structure with dislocation of oxygen anions and vacancies: *(a)* – distribution of anions and vacancies at the Ti/Al:HfO$_2$ interface after manufacturing; *(b)* – formation of a CF, i.e. SET operation; *(c)* – rupture of a CF, i.e. RESET operation.

**Fig. 2.** A cross-sectional transmission electron microscopy image of an IHP RRAM cell: *(a)* – the MIM structure with a transistor; *(b)* – the MIM structure, zoomed in.

*C. Structure of an RRAM cell*

In this work we investigate IHP RRAMs [6] manufactured in the IHP 250 nm technology [7]. Details about the switching behavior can be found in [2]. The IHP RRAM cell is based on a 1 Transistor 1 Resistor (1T-1R) architecture, i.e. an RRAM cell consists of a transistor and a MIM structure in series. **Fig. 2** shows a cross-sectional transmission electron microscopy image of an IHP RRAM cell (see **Fig. 2**–*(a)*) with the MIM structure, zoomed in (see **Fig. 2**–*(b)*).

An equivalent electrical circuit of an RRAM cell based on 1T-1R architecture is shown in **Fig. 3**. The variable resistor in **Fig. 3** represents the MIM structure in an RRAM cell.

The transistor in the IHP RRAM cell is an N-channel Metal-Oxide-Semiconductor (NMOS) field-effect transistor (see **Fig. 3**) that has two functions:

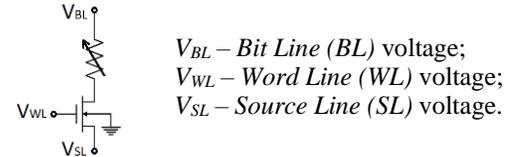

$V_{BL}$ – *Bit Line (BL) voltage*;
$V_{WL}$ – *Word Line (WL) voltage*;
$V_{SL}$ – *Source Line (SL) voltage*.

**Fig. 3.** Equivalent electrical circuit of 1 Transistor 1 Resistor architecture of an IHP RRAM cell.

- Limitation of current through the MIM structure;
- Accessibility to a cell.

Limitation of the current through the MIM structure is necessary to prevent the MIM structure from a hard breakdown. A hard breakdown occurs due to a significant change in current or voltage during electric stress. Due to this stress CFs are formed in the TMO, e.g. the HfO$_2$. It is very hard to rupture these filaments afterwards. There is also a soft breakdown that occurs by much smaller changes in current or voltage during electric stress. Due to the soft breakdown CFs are formed too, but they can be easier ruptured and recovered in the TMO.

*D. Operating modes of RRAM cells*

After fabrication the MIM-structure is in pristine state and the insulator conducts a low level of current. The resistive switching of the RRAM cells has to be initiated by a special operation called ELECTROFORMING. This operation is performed only once. ELECTROFORMING plays an important role in the subsequent performance of the RRAM cell. To write the data into the RRAM cell two operations are used: SET and RESET. The RESET operation allows to change the resistance level of the cell from the Low Resistive State to the High Resistive State (LRS → HRS). The SET operation allows to change the resistance level of the cell from the High Resistive State to the Low Resistive State (HRS → LRS). To recognize the state of the RRAM cell the READ operation is used. All the operations are performed on a special setup, see section *II-F*.

The output of the READ operation is a measured value of the current *I* that flows through the cell. The value *I* allows to determine the state of the cell. In this work we define the levels of currents that correspond to logical states as follows:

- if $(0 \leq I \leq 5)\ \mu A$ – the cell is in logical state '0' (HRS);

- if $(30 \leq I < 100)$ µA – the cell is in logical state '1' (LRS);
- if $(5 < I < 30)$ µA – the cell is in an undefined state.

Due to the fact that the manufacturing process does not allow to produce cells with ideal similarity resistive switching may not be initiated by ELECTROFORMING for some RRAM cells. Hence, these cells cannot be operated. We consider these cells as broken. Please note that the definition of the broken cells depends on the definitions of the HRS and the LRS of the cells. Additionally, it has to be taken into account that some broken cells are not really broken but only stressed and can get working over next operations. Due to these facts we used the following definition for broken cells:

- cells with $I > 1$ µA before ELECTROFORMING;
- cells with $I < 5$ µA after ELECTROFORMING or after SET operations.

Once we identified broken cells in the attacked chip we excluded these cells from the evaluation.

*E. Attacked IHP RRAM chip*

The RRAM cells in the IHP chip are organized as an array. The connection of the cells in IHP RRAM chip in an array is shown in **Fig. 4**.

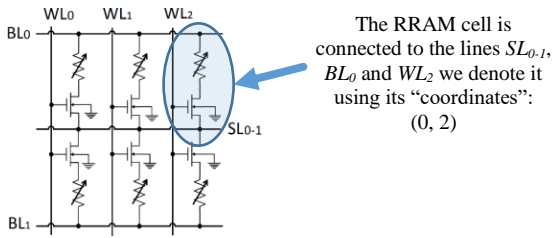

**Fig. 4.** Electrical circuit of six RRAM cells organized as an array.

Each single RRAM cell in such an array is connected to *Bit Line* $BL_i$, *Word Line* $WL_j$ and *Source Line* $SL_k$. In this work we denote cells that are connected to the $BL_i$ and $WL_j$ further as the cell with "coordinates" $(i, j)$ e.g. the cell connected to $BL_0$ and $WL_2$ is denoted further as the cell (0, 2), see **Fig. 4**. Applying voltage to a *Source Line* determines a kind of operation: RESET (if $V_{SL} \neq 0$ V) or one of the other operations (if $V_{SL} = 0$ V), i.e. ELECTROFORMING, READ or SET. The layout of *Bit Lines*, *Word Lines* and *Source Lines* will be further kept in this work as it is shown in **Fig. 4**, i.e. *Bit Lines* and *Source Lines* are horizontal lines (further rows) and *Word Lines* are vertical lines (further columns).

The arrangement of the RRAM cells as it is shown in **Fig. 4** allows to place the RRAM cells with a high density in a chip and partially accelerates the operations ELECTROFORMING, SET and RESET due to the fact that the first attempt of all these operations can be performed for all selected RRAM cells. The next attempts may be performed only to a selected row or to a single selected RRAM cell. This is due to the fact that applying the voltages again to the whole RRAM array can influence the already ELECTROFORMed/SETed/RESETed RRAM cells.

IHP manufactures chips with an RRAM array which contains 64 *Word Lines* and 64 *Bit Lines*, i.e. this array contains 64×64=4096 RRAM cells. This array can be used for storing 4 kbit of information. The layout of an IHP 4 kbit array RRAM is shown in **Fig. 5** with a part of the layout that contains only 6 RRAM cells, zoomed in.

The part of the layout with 6 RRAM cells that is shown in **Fig. 5**–*(a)* corresponds to the electrical circuit shown in **Fig. 4**. The RRAM cell marked with a circle in **Fig. 4** is marked with a rectangle in **Fig. 5**–*(a)*.

In order to control the voltages $V_{WL}$, $V_{BL}$, $V_{SL}$ and measure the current through the addressed cell the following blocks were designed at IHP [8]:

- *word* address decoder (XDC Mux).
  It is a 6-in/64-out decoder that selects a single word line $WL_i$ from the 64 word lines.
- *bit* address decoder (YDC Mux).
  It is a 6-in/64-out decoder that selects a single bit line $BL_j$ from the 64 bit lines.
- operation control circuit (Mode).

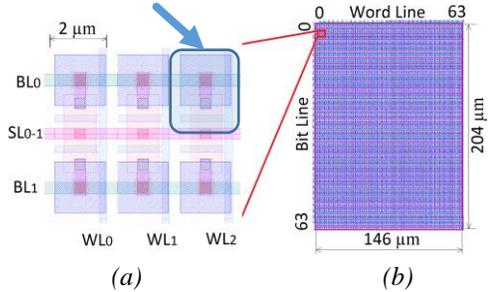

**Fig. 5.** Layout of an IHP 4 kbit RRAM chip: *(a)* – a part of the array with 6 RRAM cells, zoomed in; *(b)* – a 4 kbit memory array.

The 4 kbit memory array with these control blocks is denoted further as "4 kbit RRAM". **Fig. 6** shows a 4 kbit RRAM chip in a package with a cap (see **Fig. 6**–*(a)*) and without a cap (see **Fig. 6**–*(b)*). Details about the process of cap removal are described in section *IV*.

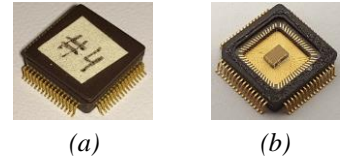

**Fig. 6.** A 4 kbit RRAM chip: *(a)* – the chip in the package; *(b)* – the chip in the package with removed cap.

IHP chips are marked with a number, e.g. chip #4 in the package, see **Fig. 6**–*(a)*.

*F. Measurement setup*

To operate with a packaged 4 kbit RRAM chip a special setup is used, see **Fig. 7**. This device is a Non Volatile Memory Tester named RIFLE SE manufactured by Active Technologies company [9]. **Fig. 7** shows the RIFLE SE operating device (see **Fig. 7**–*(a)*) and a PCB with a socket for a 4 kbit RRAM chip (see **Fig. 7**–*(b)*), zoomed in.

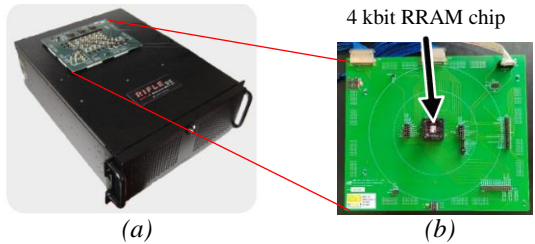

**Fig. 7.** Operating device for a 4 kbit RRAM chip:
*(a)* – Active Technologies RIFLE SE;
*(b)* – a PCB with a socket for a 4 kbit RRAM chip.

Due to the big size of the RIFLE SE device (90 cm × 75 cm) it is impossible to place it into the box with the laser. This limits the experiments that can be performed, e.g. the influence of the laser during the READ, ELECTROFORMING, RESET and SET operations cannot be measured. Therefore we first measured the currents through the RRAM cells, i.e. we performed the READ operation for each cell, after that we performed the laser scans and then we measured the currents through the cells again to compare the values of $I$ before and after the laser scan. The output of the performed READ operation using the RIFLE SE device is a text file (*.txt). This file contains 64×64 values of $I$ through each of the 4096 cells, i.e. it is a two-dimensional array corresponding to the RRAM cells topology. We denote this array further as matrix of Measured Currents (*MC*). Each element in an *MC*-matrix is in the range from 0 up to 100 µA, see section *II-D*.

### III. OPTICAL FI BACKGROUND AND FI SETUP

#### A. Basics: exploiting of photoelectric effect for FI

A photoelectric effect is observed in a variety of materials such as metals, semiconductors and dielectrics. There are two types of the photoelectric effect:
- external photoelectric effect; it is observed in metals;
- internal photoelectric effect; it is observed in semiconductors and dielectrics.

Details about the external photoelectric effect can be found in [10]. We describe here shortly the internal photoelectric effect because the attacked cells contain an active oxide layer (dielectric) used in a MIM structure that is responsible for the filament forming and an access transistor (semiconductor device). In semiconductors electrons are "located" in the atom orbitals. Electrons can absorb energy of the laser irradiation and leave their atoms if the absorbed energy is sufficient. This leads to an appearance of the free electrons – the photoelectrons – as well as their holes in the material. Subsequently many photoelectrons can increase the conductivity of a material significantly. Hence, using a laser irradiation precisely it is possible, e.g. to switch a selected transistor from a high resistance state (closed) to low resistance state (open).

The goal of our experiments is to investigate the sensitivity of IHP RRAM cells to optical FIs. Due to the known sensitivity of the MIM structures to heating [11] we expect that the laser irradiation, especially infrared laser, can influence the RRAM cell by changing the value of $I$ through the MIM structure.

Accordingly to the process of the CFs formation/rupture we assume that the illumination of the MIM structure(s) leads to a possible additional creation/recombination of the Frenkel pairs at the Ti/Al:HfO$_2$ interface, especially if the laser beam hits the MIM structure. This creation/recombination of the Frenkel pairs leads to thickening/thinning or formation/rupture of the CF that influences the conductivity of the MIM structure. It is not known which process – the formation or the rupture of the CF – will be triggered when the laser illuminates/hits the MIM structure.

#### B. Fault Injection Setup

Our Fault Injection setup is shown in **Fig. 8**. It comprises:
- A PC with the Riscure Inspector FI software [12]. It allows to create a so-called "FI program". An FI Program allows to store the applied set of parameters in an experiment.
- A Riscure VC glitcher [13]. This device is the block that generates faults (see the block in **Fig. 8** placed between the block PC and the block Laser).
- The laser source. It generates light pulses corresponding to the set of parameters saved as an FI Program.
- The optical system. It allows to focus the beam and reduce the spot size to micrometer units.
- An X-Y stage. It allows to move the attacked chip and perform scanning.

The PC is connected with an X-Y stage, microscope camera and a VC glitcher via USB. The VC glitcher in turn is connected with a laser source via a cable with SMA connectors. All blocks except PC, light source and X-Y control unit with joystick are placed into a Safety box. It protects the user from possibly harmful reflections of the laser beam.

We used in our experiments the 1$^{st}$ generation Riscure Diode Laser Station (DLS) [14], see **Fig. 8**–*(b)*. Corresponding to the Riscure data sheet [14] the parameters of the DLS are:
− The DLS is equipped with two multimode laser sources:
  ○ 14 W for the red laser 808 nm;
  ○ 20 W for the near infrared (NIR) laser 1064 nm;
− pulse duration in a range of 20 ns – 100 µs;
− trigger delay is 50 ns;
− elliptical spot sizes 60×14 µm$^2$, 15×3.5 µm$^2$, 6×1.5 µm$^2$ and 3×0.75 µm$^2$;
− magnification objectives: 5×, 20×, 50×, 100×;
− Filters: 10%, 1%, 0.1%;
− X-Y stage with 3 µm accuracy and 0.05 µm [14] step distance between two adjacent points.

The filter in the setup is used to reduce the laser beam output power. Usually the NIR laser source is used to perform experiments through the back-side of the targeted chip due to the fact that the substrate is transparent to the infrared spectrum [15]. We performed attacks using the red laser source through the front-side, but we assume that a utilization

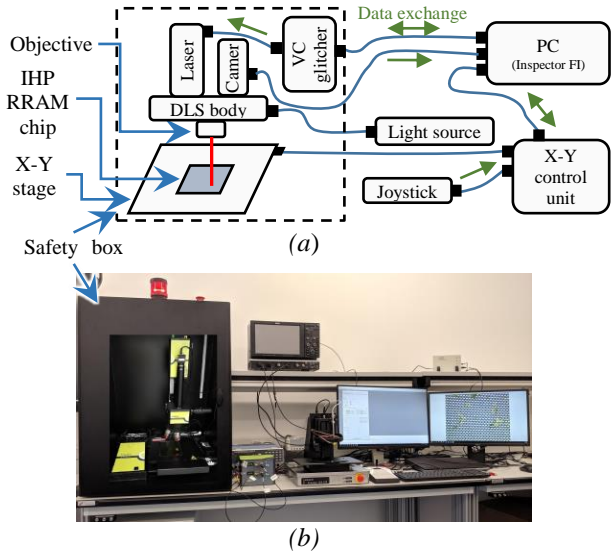

**Fig. 8.** Fault Injection setup: *(a)* – schematic view; *(b)* – setup at IHP.

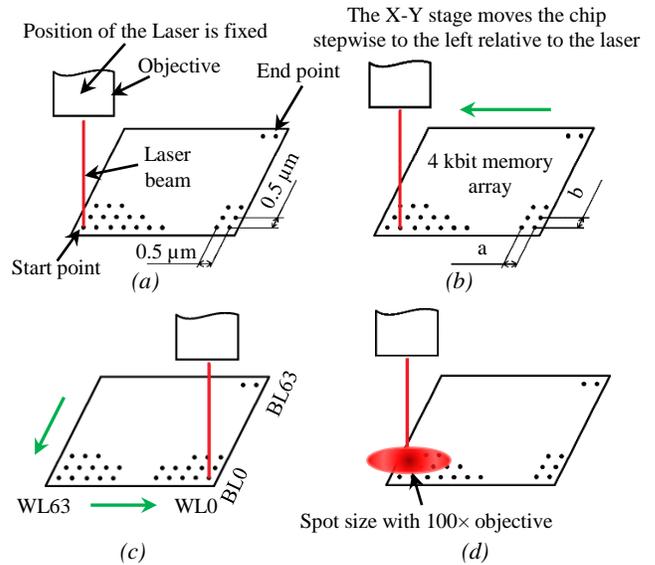

**Fig. 9.** Position of the attacked chip on the X-Y stage:
*(a)* – 1st (start) point in the 1st *X*-movement line; *(b)* – 2nd point in the 1st *X*-movement line; *(c)* – last point in the 1st *X*-movement line; *(d)* – 1st point of the 2nd *X*-movement line.

The distance between two adjacent points in each *X*-movement line is 0.5 μm for the *X* axis; the distance between two neighbor *X*-movement lines is 0.5 μm for the *Y* axis. After the first *X*-movement line is scanned the table moves to the start point, and after this moves along the *Y* axis to the next *X*-movement line.

of NIR laser will be more effective, see section *III-A*. We use the red laser because it has a lower laser beam output power, i.e. a lower probability to destroy the chip.

The position of the Laser as a device is fixed in the Riscure set-up. The attacked chip has to be moved with the goal to illuminate its different parts. The active movement is performed by X-Y stage manufactured by Märzhäuser Wetzlar GmbH & Co [15]. Due to the fact that the step of the X-Y stage that is given in the Riscure datasheet is too small and equal to 0.05 μm that is 50 nm, we performed some measurements to verify the step size. The result of our measurements is: the step distance between two neighboring points is 0.25 μm. If the programmed step distance was 0.2 μm or less we noticed some mismatches between the number of steps programmed using the Riscure Inspector FI software and the number of steps done when observing through the microscope camera. Hence, the range and the minimal distance of movements are corresponding to our measurements as follows:

- Along the X axis the range of movement is 75 mm with a minimum possible distance between two adjacent points of 0.25 μm;
- Along the Y axis the range of movement is 50 mm with a minimum possible distance between two adjacent points of 0.25 μm.

In our experiments we selected the distance of 0.5 μm between two adjacent points due to the area occupied by a single MIM structure (0.6×0.6 μm$^2$) in the attacked IHP RRAM chips. **Fig. 9** shows the position of the stage in our experiments with the selected step distance between two neighboring points.

Please note that the area of the laser beam spot depends on the objective used in our experiments. Due to the selected step distance between two neighboring points, the size of the RRAM cell (2.2×3 μm$^2$) and the area of the laser beam spot a different number of the RRAM cells can be illuminated by one laser shot.

## IV. PREPARATION OF THE CHIP FOR THE EXPERIMENTS

In total we performed scans with three new 4 kbit RRAM chips. The attacked chips are marked with numbers #2, #4 and #9. The acquired 4 kbit RRAM chips were in packages, see **Fig. 6**–*(a)*. In order to perform laser experiments the access to the chip surface is needed. Hence, the cap of the package was removed with a hot air work station for PCBs. The decapped chip is shown in **Fig. 6**–*(b)*. For successfully removing the cap we heated the chip 15 s with a temperature of 450 degrees Celsius. After the cap removal we made a visual inspection to check if the bonding wires were still intact and the chip surface is clean. After the chip had been tested for functionality it was programmed.

### A. Programming of the 4 kbit RRAM chips

The cells in two attacked chips were programmed with different logical states. This was done in order to assess how the laser influences the cells in a different states.

#### Chip #4 and Chip #9

After the ELECTROFORMING the chips #4 and #9 were programmed to the logical states '1' and '0' as follows:

- cells with *Word Lines* 0-15 – block 1 – were programmed to the logical state '0'.
- cells with *Word Lines* 16-31 – block 2 – were programmed to the logical '1'.
- cells with *Word Lines* 32-47 – block 3 – were programmed to the logical '0'.
- cells with *Word Lines* 48-63 – block 4 – were programmed to the logical '1'.

**Chip #2**

After the ELECTROFORMING of the chip #2 the logical state '1' was stored in each cell of the entire 4 kbit memory array.

TABLE I shows the result of storing data in 4096 cells of each of the 3 investigated chips.

TABLE I. OVERVIEW OF THE RESULTS OF THE STORING OF DATA

| Chip Nr. | '0' | '1' | Undefined state | Broken cells |
|---|---|---|---|---|
| #4 | 1765 | 1770 | 62 | 499 |
| #9 | 0 | 969 | 27 | 3100 |
| #2 | 0 | 2613 | 530 | 953 |

After storing the data we placed the chip on the X-Y stage.

Please note that as a means to determine the measurable influence of laser irradiation on the RRAM cells we performed READ operations many times with different time intervals between the READ operations: 2 minutes, 1 day and 1 week. We measured 25 $MC$-matrices for different chips and analyzed them statistically. The result of the analysis shows that 99.450 % of changes of the measured current $\Delta I$ between different $MC$-matrices for a single chip are in the range $|\Delta I| < 4.084$ µA. Hence, if the current changed more than $|\Delta I| \geq 4.084$ µA we consider it further as a successful laser influence.

## V. LASER EXPERIMENTS AND DISCUSSION OF THE RESULTS

The goal of our experiments was to define the parameters of the laser set-up for a successful and repeatable FI without any knowledge about the sensitivity zones in RRAM structures. Due to the fact that no voltages are applied to the cells in the attacked RRAM chip the injection of faults into the NMOS transistor is not possible. However we scanned the whole 4 kbit memory array because the IHP technology requires the implementation of the metal fillers in manufactured chips. The metal fillers are relatively small metal areas that are placed in different metal layers between the wires to maintain the chip stiffness. Hence, the wires inside the chip and metal fillers are obstacles to the laser beam and can reduce the success of the FIs. We started with the chip #2. At first we performed the READ operation. The measured currents were stored as matrix $MC_0^{\#2}$. We denote here as an experiment a laser scan of the 4 kbit RRAM memory array that we performed using a certain set of parameters. The first laser scan we performed with the following parameters: 50% power; 20 ns pulse duration; 10% filter; 1 shot per move; 0.5 µm distance between two adjacent cells for both X- and Y-axis; 100× objective. We began with the minimal laser pulse duration and low energy to avoid harming the analyzed chip. After the scan we performed the READ operation; the measured currents were stored as the matrix $MC_1^{\#2}$. To evaluate the success of the FIs in the experiment we calculate the difference of the matrices $\Delta_{m,k}^{\#2}$, e.g. $\Delta_{1,0}^{\#2} = MC_1^{\#2} - MC_0^{\#2}$ and analyzed it. Each element $\delta_{i,j}$ of the matrix $\Delta_{m,k}^{\#2}$ is a difference of the measured currents $I$ of the cells with coordinates ($i,j$) of the matrices $MC_m^{\#2}$ and $MC_k^{\#2}$: $\delta_{i,j}^{\#2} = (I_{i,j})_m^{\#2} - (I_{i,j})_k^{\#2}$. If a value $\delta_{i,j}^{\#n}$ is changed by more than 4.084 µA, i.e. if $|\delta_{i,j}^{\#n}| \geq 4.084$ µA, we consider it as an observable influence of the laser. If the influenced cell changed its logical state, we consider it as a Fault Injection, see section *IV-A*. Hence, for each value $|\delta_{i,j}^{\#n}| \geq 4.084$ µA in the Δ-matrix we check the logical state of the cell with the coordinates ($i, j$) before and after the performed laser scan. TABLE II shows the results of the first FI experiment. The laser influence was observable for 250 cells but only for 72 cells FIs were observable, i.e. the laser influence caused a change in the logical state of the cells. **Fig. 10** shows a matrix that visualizes the successful FIs in RRAM cells after the first experiment with chip #2, i.e. it represents the data shown in TABLE II graphically. The color legends in TABLE II and **Fig. 10** are the same.

TABLE II. OVERVIEW OF THE SUCCESSFUL FAULT INJECTIONS FOR THE 1ST LASER SCAN EXPERIMENT WITH THE CHIP #2.

| State of the cells | | | | | | |
|---|---|---|---|---|---|---|
| Unchanged state | Bit-reset '1'→'0' | Bit-set '0'→'1' | '0'→undef. | '1'→undef. | undef.→'0' | undef.→'1' |
| 3071 | 8 | 0 | 0 | 21 | 31 | 12 |

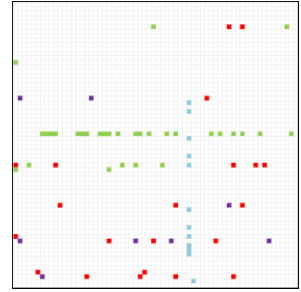

**Fig. 10.** The visualized matrix of the successful FIs of the first experiment with the chip #2.

The RRAM cell with "coordinates" (0, 0) is shown in the top left corner of the matrices in **Fig. 10**. The bottom right corner of the matrices corresponds to the RRAM cell with coordinates (63, 63). In the following experiments we increased the power, pulse duration, number of shots per move and applied different objectives. The parameter set for each scan-experiment is given in TABLE III. After each scan-experiment $l$ we performed the READ operation, stored the matrix of measured currents $MC_l^{\#2}$, and evaluated the success of the FIs calculating the difference matrix $\Delta_{l,l-1}^{\#2} = MC_l^{\#2} - MC_{l-1}^{\#2}$ applying the threshold $|\delta_{i,j}^{\#n}| \geq 4.084$ µA as the criterion for measurable laser influence.

We performed similar experiments and analysis for the chip #4 and for the chip #9. TABLE III gives the results of these experiments.

Please note that the values of measured currents $I$ for any two READ operations are always slightly different. Due to this fact for some cells their logical state can change even without laser illumination (but only undefined state ↔ '1' or undefined state ↔ '0' are possible). In TABLE III we listed only the cases that were identified as measurable laser influence, i.e. only cases with $|\delta_{i,j}^{\#n}| \geq 4.084$ µA.

TABLE III. RESULTS OF LASER SCAN EXPERIMENTS

| Chip Nr. | Experiment | Automated (a)/manual (m) | Parameters of DLS | | | | | Laser influence | | | | | | | |
|---|---|---|---|---|---|---|---|---|---|---|---|---|---|---|---|
| | | | | | | | | FI | | | | | | no change in logic state of cells | |
| | | | Power, % | Pulse duration, ns | Objective | Filter, % | Shots per move | '0'→'1' | '1'→'0' | '0'→undefined | '1'→undefined | undefined→'0' | undefined→'1' | '0'→'0' | '1'→'1' | undefined→undefined |
| #2 | 1 | a | 50 | 20 | 100× | 10 | 1 | 0 | 8 | 0 | 21 | 31 | 12 | 0 | 165 | 13 |
| | 2 | | 70 | 20 | 100× | | 1 | 9 | 0 | 9 | 10 | 0 | 16 | 1 | 132 | 6 |
| | 3 | | 100 | 100 | 100× | | 1 | 0 | 0 | 5 | 0 | 0 | 5 | 0 | 34 | 5 |
| | 4 | | 100 | 10⁵ | 100× | | 1 | 14 | 0 | 2 | 1 | 0 | 31 | 0 | 38 | 5 |
| | 5 | | 100 | 10⁵ | 50× | | 1 | 0 | 0 | 0 | 3 | 0 | 8 | 0 | 40 | 7 |
| | 6 | | 100 | 10⁵ | 20× | – | 3 | 0 | 0 | 0 | 2 | 0 | 11 | 0 | 79 | 6 |
| | 7 | | 100 | 10⁵ | 5× | | 3 | 0 | 0 | 0 | 1 | 0 | 3 | 0 | 14 | 0 |
| | 8 | m | 100 | 5·10⁶ | 100× | | 1-5 | 0 | 0 | 0 | 1 | 0 | 0 | 0 | 0 | 0 |
| | 9 | | | | 20× | | | 0 | 0 | 0 | 7 | 0 | 0 | 0 | 4 | 2 |
| | 10 | | | | 5× | | | 0 | 5 | 0 | 27 | 1 | 1 | 0 | 25 | 5 |
| #4 | 1 | a | 100 | 40 | 100× | | 1 | 0 | 0 | 18 | 0 | 2 | 0 | 8 | 3 | 0 |
| | 2 | | 100 | 10⁴ | 100× | | 1 | 0 | 0 | 6 | 0 | 0 | 0 | 1 | 1 | 0 |
| | 3 | | 100 | 10⁵ | 100× | | 3 | 0 | 0 | 1 | 0 | 2 | 1 | 0 | 2 | 0 |
| | 4 | | | | 100× | – | | 0 | 0 | 1 | 0 | 0 | 0 | 0 | 0 | 0 |
| | 5 | | 100 | 10⁶ | 20× | | 1-5 | 0 | 0 | 5 | 1 | 1 | 0 | 1 | 1 | 0 |
| | 6 | m | | | 5× | | | 0 | 0 | 7 | 0 | 2 | 0 | 1 | 1 | 0 |
| | 7 | | | | 100× | | | 0 | 0 | 0 | 0 | 0 | 0 | 0 | 0 | 0 |
| | 8 | | 100 | 5·10⁷ | 20× | | 1-5 | 0 | 298 | 3 | 13 | 63 | 5 | 49 | 113 | 1 |
| | 9 | | | | 5× | | | 10 | 584 | 59 | 16 | 20 | 2 | 0 | 57 | 4 |
| #9 | 1 | a | 50 | 20 | | | 1 | 0 | 0 | 0 | 0 | 0 | 7 | 0 | 53 | 2 |
| | 2 | | 90 | 20 | 100× | – | 1 | 0 | 0 | 0 | 0 | 0 | 1 | 0 | 26 | 1 |
| | 3 | m | 100 | 10⁷ | | | 1-5 | 0 | 975 | 0 | 0 | 17 | 0 | 1 | 0 | 0 |

Not all experiments that we performed were done in the automated mode. In many experiments the pulse duration and number of shots per move were adjusted manually, see TABLE III. The automated mode is limited by the 100 µs pulse duration but it allows to program the X-Y stage and to perform the scanning of the attacked chip automatically, always with the same – programmed and stored – parameters: laser beam output power, pulse duration, movement distance, shots per move, etc. In the manual mode the X-Y positioning stage will be controlled manually using a joystick (manipulator), see **Fig. 8**. This mode allows to bypass the limitation of the 100 µs for the maximum pulse duration, but every shot should be performed manually. The pulse durations we applied in our experiments were calculated for the "Glitch Source" tab in "SC Single XYZ Perturbation" menu[1]. Due to the longer pulse duration than in the automated mode there is a high probability that the surface of the chip or the chip itself may be damaged, e.g. RRAM cells, decoders, operation control circuit. For this reason each of three attacked IHP RRAM chips was scanned in an automated mode, first.

**Fig. 11** shows a total number of influenced RRAM cells for each chip in all experiments conducted, i.e. it represents the data given in TABLE III graphically.

---

[1] in Riscure FI Inspector software, SC Single XYZ Perturbation → Glitch Source: pulse duration = "Clock speed" · · "Test glitch source cycles".

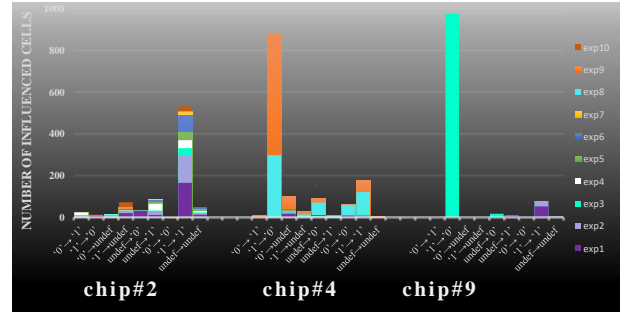

**Fig. 11.** Total number of influenced RRAM cells for each chip in all experiments conducted.

**Experiments with Chip #2**, see TABLE III:
- successful FIs in automated mode are more likely with low laser beam power and short pulse duration;
- successful FIs in manual mode are more likely with long pulse duration and use of small magnification objective.

The surface of chip #2 after the experiments in manual mode was damaged, i.e. it shows visible traces of the laser irradiation. However the internal structure of the chip was not damaged. Chip #2 is still fully functional after all experiments performed, i.e. it still responds to the READ operation without any errors.

**Experiments with Chip #4**, see TABLE III:
- successful FIs in automated mode are more likely with short pulse duration;
- successful FIs in manual mode are more likely with long pulse duration and use of small magnification objective.

Chip #4 was damaged during the experiments. The surface (passivation layer) of chip #4 was damaged after the experiments 4-6. However the chip continued functioning, so we assume that its internal structure was still intact. After experiment 9 the chip #4 stopped functioning and could not be operated anymore. Hence, we assume that changes of the cell states given in the experiment 9, see TABLE III occurred because of a damage of the internal structure. It means that a pulse duration of 50 ms and the use of 5× magnification objective can damage not only the surface but internal structures of the chip, also.

**Experiments with Chip #9**, see TABLE III:
- successful FIs in automated mode are more likely with low laser beam power.

We performed only a small number of experiments with chip #9 because it was mechanically damaged after the 3rd experiment, i.e. several bonding wires were detached from the chip pads during transportation of the chip to the measurement equipment, see subsection *II-F*. We do not know exactly if the set of parameters used in experiment 3 caused any damage to the internal structure of the chip but we saw the damage of the chip surface through the microscope camera.

According to the visual inspections of the chips before and after the experiments we can conclude that the configuration

of the variable parameters in our experiments:
- in the automated mode leads to no damage, neither to the passivation layer nor to the internal structure;
- in the manual mode may lead to the distinct visible damage of the chip surface with the following parameters: 1 ms pulse duration; 100% laser output power; 100× objective;
- in the manual mode a damage of the chip internal structure may occur with the following parameters: 50 ms pulse duration; 100% laser output power; 5× objective.

Comparing the data in TABLE III we can conclude that the more successful FIs in automated mode are observed with the following parameters:
- a low laser beam output power;
- a short pulse duration;
- an application of the 100× magnification objective.

Experiments in a manual mode show that:
- the success of FIs is more likely with a long pulse duration and the use of small magnification objectives;
- the damage of the surface and the damage of the internal structure of the chip is also more likely with a long pulse duration and the use of small magnification objectives.

Generally results of our experiments are:
- the change of the logical states as the result of the laser influence is observable in all attacked chips;
- reaction of chips to laser illumination is individual (more RRAM cells changed their logical state for chip #4);
- we are not sure if the reduced number of the cells influenced by the laser in sequentially executed experiment is caused by the new parameter set only (it has to be taken into account that cells that are easy to influence probably switched their logical state already in the previous experiment).

Nevertheless the sets of parameters applied in our laser experiments that cause successful FIs require additional experiments to be confirmed. Thus, the experiments with low laser beam output power should be performed in order to confirm or disprove the influence of the laser on the cells state.

## VI. CONCLUSION

In this work we investigated the sensitivity of the IHP RRAM chips to optical Fault Injection attacks. We demonstrated that laser based attacks against RRAM chips can be successful. Lasers can influence the state of the cell significantly, i.e. cells can change their logical state. Changes of cell states were observed for all attacked IHP RRAM chips. The results show that the cells are prone to change their states with the selection of the following laser irradiation parameters: low laser beam power, short pulse duration and the use of the high magnification objectives. In this case the changes of the cells states are achieved without any visible traces of the laser irradiation on the chip surface and no damage to the internal structure. The influence of the laser irradiation on the cells states vanishes with the increase of these three parameters. Nevertheless with a significant increase of the values of the laser parameters, i.e. laser beam power, pulse duration, etc., we observed successfully injected faults again. However the increase of the values of these parameters caused damage of the surface and may be accompanied with a damage of the internal chip structure.

ACKNOWLEDGMENT

This project has received funding from the European Union's Horizon 2020 research and innovation program under the Marie Skłodowska-Curie grant agreement No 722325.